# Identification of similarity amongst eucalyptus planted areas based on leaf-cutting ant nest sizes

D M L Barbato[1], J M A De Andrade[2], J J De Groote[3]

[1] Instituto SEB de Educação, Rua Deolinda 70, Jardim Macedo, CEP 14091-018, Ribeirão Preto, São Paulo, Brazil

[2] Centro Universitário UNISEB|ESTÁCIO, Rua Abraão Issa Hallack 980, CEP 14096-175, Ribeirão Preto, São Paulo, Brazil

[3] Centro Universitário Moura Lacerda, Av. Dr. Oscar de Moura Lacerda, 1520, CEP 14076-510, Ribeirão Preto, São Paulo, Brazil

**Abstract**

Techniques for leaf-cutting ant control have been investigated in literature due to the importance of the damage they cause to agriculture. The effectiveness of different forms of control is explored in researches aimed at identifying the balance between pest control efficiency and environmental damage caused by the treatments applied. Plantations with large territorial extensions, which can be contiguous or not, are usually subdivided into local administration that collects data to determine the frequencies of areas size occupied by ant nests. The purpose of this work is to build a relationship of similarities among different geographical regions using the frequency data of nests size occurrence by applying Information Bottleneck (IB) method and Principal Component Analysis (PCA). IB allows simultaneous clustering of each region with the ant nest size distribution, while PCA is used to reduce the variable dimensionalities into a three-dimensional representation of the results. The approach was applied to data of leaf-cutting ants *Atta* spp. (Hymenoptera: Formicidae) in cultivated *Eucalyptus* spp forests in São Paulo, Brazil. The results suggest information acquired by the method can help coordinate pest management, such as the allocation of baits, material and personnel.

**Keywords:** Information bottleneck, Leaf-cutting ant, Pest management effectiveness, Planted forest pest management, spatial patterns of ant colonies.

**Introduction**

Leaf-cutting ants are important pests of several crops, pastures and planted forests, such as commercial Eucalyptus forest. Ants belonging to the *Attini* tribe live in the Americas, from Southern United States to Argentina center in South America (Pérez et al. 2011) Brazil has the largest number of leaf-cutting ant species, classified into two kinds, *Acromyrmex Mayr* and *Atta Fabricius (Hymenoptera: Formicidae)*. Both attack cultivated plants and use the vegetables to nurture a fungal culture they feed on (Zanetti et al. 2014).

Damage caused by the ants are relevant since they continuously cut tender leaves and branches of plants at any stage of development, leading to intense defoliation which can even destroy them altogether (Souza et al. 2011).

For planted forests, they are considered the worst pests, being responsible for significant losses, with expenses in their control reaching up to 30% of forest cost by the end of the third cutting cycle.

In Brazil, planted forests are mainly (93%) of pine and eucalyptus, totalizing 7.3 million hectares in 2016, of which 78.1% are *Eucalyptus* L'Hér. (Myrtaceae) and 21.9%, Pinus L. (Pinaceae) (IBÁ 2017). The highest concentration of eucalyptus and pine plantations occurs in the South and Southeast of Brazil (72%), where the niche market of their products is located. Regarding the eucalyptus, the economic interest is mainly due to paper and cellulose industry that concentrate 72.5% in planted area, followed by steel and coal (19.5%), wood panels (7.3%) and independent producers (0.7%) (ABRAF 2013).

Economic loss caused by leaf-cutting ants in crops have led to investigations of efficiency and impact of different insecticides (Oi et al. 2000, Choate et al. 2013), new baits carriers (Buczkowski et al. 2014) and ant colony mapping to understand and predict their potential expansion (Morrison et al. 2004, Pitt et al. 2009).

The importance of insect control has also encouraged researches focused on statistical modeling of the spatial distribution of nests (Doncaster 1981, Croft et al. 1983, Caldeira et al. 2005, Nickele et al. 2010, Gao 2013, Sousa-Souto et al. 2013). It allows the development of better ant bait distribution strategies through more appropriate sampling plans than common basic standardization applied in most Brazilian plantations (Nickele et al. 2013, Bollazzi et al. 2014).

The occurrence of the leaf-cutting ant nests can be related to several environmental factors such as sunlight exposure, humidity, altitude and food availability (Doncaster 1981, Kaspari et al. 2000). To a lesser degree, dispersion methods and competitive strategies may be considered (Tanner et al. 2012, Ricklefs et al. 1994).

In terms of insect spatial arrangements there are three main discrete statistical models, defined as random, described by Poisson distributions, contagious, negative binomial distributions, and regular or uniform, positive binomial distributions (Taylor 1984; Begon et al. 1996). By modeling the distribution, it is expected that better planning helps to decrease the use of bait, avoiding considerable expenses and also control deleterious effects to the environment resulting from excessive use of insecticides (Laranjeiro 1994, Nickele et al. 2013).

Plantations with large territorial extensions, which can be contiguous or not, are usually subdivided into local administration (geographical sub-regions) that collects data to determine the frequencies of area size occupied by the ant nests.

For such cases, identification of similarities among different geographical sub-regions based on nest distribution can also be relevant information to determine cost reduction strategies by improving resource quantification to reduce loss and also avoid unnecessary environmental contamination.

Standard procedure for determining nest size consists in dividing regions into plots (stands) that are subdivided to identify and estimate the nest area using the greatest length and width (Caldeira et al. 2005).

In cases of great variability in the sizes of nests, the amount of data collected and the number of nest size intervals can be large, thus, making it difficult to infer the correlation between regions. For such cases, it becomes necessary to investigate statistical methods developed for multivariate data analysis.

Different data clustering algorithms are used in information retrieval problems, image segmentation, pattern classification, phylogenetic inference, microarray gene expression, and for most of them, it is necessary to pre-set parameters as the distance between pairs of points or distortion rate between point and class. These parameters are arbitrary and influence on the result. For certain problems, the definition of an appropriate measure is a difficult task. Information Bottleneck is an unsupervised clustering method based on Information Theory, which allows information compression while preserving its relevance.

In order to improve results analysis the IB method was applied in conjunction with the Principal Component Analysis, which is used for multivariate data analysis to uncorrelated variables, and is useful as a tool to reduce the dimension size to facilitate data structure analysis.

In the next section a description of the techniques applied are presented, followed by results and discussions.

**Material and Methods**

*Principal Component Analysis*

Principal Component Analysis is a method used in the study of large multidimensional dataset. This method is based on the covariance matrix and their eigenvalues and eigenvectors determination. The eigenvector with bigger eigenvalue is principal component, related to the largest variance. Original data are projected into orthogonal eigenvector base, in such a way that they become uncorrelated. The dimension can be reduced by choosing eigenvectors with bigger eigenvalues, which explains the largest variance of the sample (Jolliffe 2011). In the new base, we can observe clusters, but in some cases, its configuration is not obvious, therefore other methods are necessary.

PCA in this work is an auxiliary tool to reduce the data variables dimension allowing the visualization of results produced by IB method.

*Information Bottleneck*

Given the *X* and *Y* variables and their joint probability distribution *p*(*x*, *y*), the variable *X* must be compressed in order to maintain maximum information about variable *Y*. The degree of compression is represented by a variable *T* and indicates the number of clusters formed. If the compression is too big, there is information loss, but if there is no compression, analysis becomes more difficult. This method consists the minimization of a functional based on the mutual information between *X* and *T*, $I(X,T)$, representing compression, and between *Y* and *T*, $I(Y,T)$, which represents relevancy. A functional parameter *β* (inverse temperature) mediates compression and relevance.

IB method has been applied in different areas such as speech recognition (Hecht et al. 2009), galaxy spectra classification (Slonim et al. 2001), psychometric properties of questionnaires (Barbato & De Groote 2018) and unsupervised document clustering, essential for problems information retrieval (Slonim & Tishby 2000), where similarity measure between documents is the similarity between their conditional distributions of words. Variants for geometric clustering have also been developed (Strouse & Schwab 2019).

We search a compressed representation of *X*, through the variable *T*, which preserves a significant fraction of information about *Y*.

IB method aims to minimize the functional

$$L[p(t|x)] = I(X,T) - \beta I(T,Y), \quad (1)$$

where $I(X,T)$ is mutual information between variables $X$ and $T$, and $I(Y,T)$ is mutual information between variables $Y$ and $T$. The Lagrange multiplier $\beta$ controls tradeoff between compression and relevance.

The first term in Eq. (1) is related to compression and the second is related to relevance. For $\beta \to 0$ we have maximum compression and information loss, for $\beta \to \infty$ only relevance is preserved. For finite $\beta$, we can find a compressed $X$ representation without significant loss of relevance. For given $T$, $\beta$ and $p(x,y)$, IB method provides a probability distribution $p(t|x)$ that minimizes the functional $L$.

By minimizing the functional $L$, we find conditional probabilities $p(t|x)$ of an $L$ local minimum.

The IB method is characterized by the iterative algorithm,

- Input: $p(x,y), \beta$ e $T$.
- Initialization: $p(t|x)$ is set with random values.
- Iterative equations,

$$p(t_i) = \sum_j p(x_j) p(t_i | x_j) \quad (2)$$

$$p(y_i | t_j) = \sum_k p(y_i | x_k) p(x_k | t_j) \quad (3)$$

$$p(t_i | x_j) = \frac{p(t_i)}{Z(x_j, \beta)} EXP\left(-\beta \sum_k p(y_k | x_j) \log(p(y_k | x_j) / p(y_k | t_i))\right), \quad (4)$$

where $Z$ is the partition function defined as,

$$Z(x_j, \beta) = \sum_i p(t_i) EXP\left(-\beta \sum_k p(y_k | x_j) \log(p(y_k | x_j) / p(y_k | t_i))\right) \quad (5)$$

This algorithm finds the probability distributions $p(t|x)$ and $p(y|t)$, that shows how variables are grouped, i.e., which variables belong to each cluster.

In this work, $X$ is the range of the histogram, which represents nest area, $Y$ represents regions and $T$ represents the number of clusters. We aim to find how regions can be grouped according to size nest occurrence. For fixed $\beta$ and $T$ we find the best cluster configuration, as $\beta$ grows up, the probability $p(t|x) \to 1$.

We used the ant nests data presented by Andrade et al. (2014), obtained from seven regions of eucalyptus plantations in São Paulo State, $\{R_i, i=1..7\}$, located as shown in Fig. 1. The regions begin in the Itararé town to São Simão, comprising about 43,000 hectares of plantations. The proposal is to identify how data obtained from these regions are correlated. These data were drawn from a specific database obtained by local administration of each study area (Andrade et al. 2014).

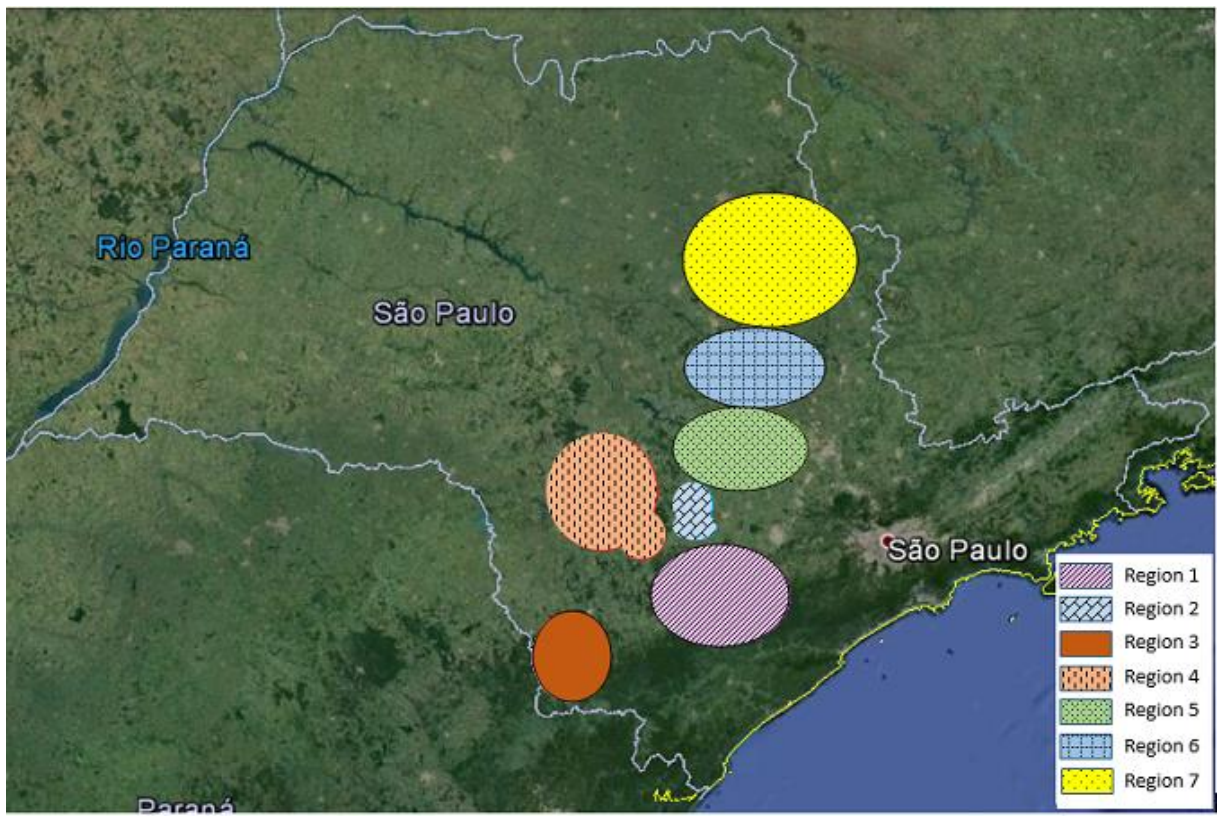

Fig. 1. Regional distribution of ant nests indicating where data was collected.

The average size of the nests found in each region was calculated to determine the occurrence frequency according to classes ranged in group intervals of 5 m² from 0 m² to mean areas equal to or greater than 300 m², as seen in Table 1.

Table 1. Nests occurrence frequency from different regions ($R_i$) with data collected in 2012 (Andrade et al. 2014). Area values without nests occurrence are not shown.

| Nests area (m²) | $R_1$ | $R_2$ | $R_3$ | $R_4$ | $R_5$ | $R_6$ | $R_7$ |
| --- | --- | --- | --- | --- | --- | --- | --- |
| 0-5 | 167 | 60 | 146 | 55 | 145 | 91 | 62 |
| 5-10 | 4 | 18 | 4 | 24 | 10 | 13 | 41 |
| 10-15 | 0 | 12 | 3 | 16 | 5 | 9 | 32 |
| 15-20 | 1 | 5 | 2 | 16 | 3 | 8 | 26 |
| 20-25 | 0 | 11 | 3 | 2 | 3 | 2 | 15 |
| 25-30 | 1 | 6 | 1 | 9 | 2 | 7 | 7 |
| 30-35 | 1 | 8 | 0 | 9 | 4 | 4 | 0 |
| 35-40 | 2 | 5 | 4 | 9 | 1 | 2 | 0 |
| 40-45 | 1 | 4 | 0 | 5 | 1 | 4 | 0 |
| 45-50 | 0 | 6 | 0 | 9 | 3 | 3 | 0 |
| 50-55 | 0 | 4 | 1 | 5 | 0 | 5 | 0 |
| 55-60 | 1 | 9 | 2 | 3 | 0 | 4 | 0 |
| 60-65 | 0 | 2 | 0 | 4 | 0 | 3 | 0 |
| 65-70 | 1 | 1 | 2 | 1 | 1 | 1 | 0 |
| 70-75 | 1 | 4 | 1 | 1 | 1 | 5 | 0 |
| 75-80 | 0 | 1 | 3 | 0 | 0 | 2 | 0 |
| 80-85 | 0 | 2 | 0 | 0 | 0 | 1 | 0 |
| 85-90 | 1 | 4 | 0 | 2 | 1 | 4 | 0 |
| 90-95 | 0 | 1 | 0 | 2 | 0 | 1 | 0 |
| 95-100 | 0 | 1 | 0 | 3 | 0 | 2 | 0 |
| 100-105 | 1 | 0 | 0 | 1 | 0 | 1 | 0 |
| 105-110 | 0 | 1 | 0 | 0 | 0 | 4 | 0 |
| 110-115 | 0 | 2 | 0 | 1 | 0 | 0 | 0 |
| 115-120 | 0 | 3 | 2 | 0 | 0 | 0 | 0 |
| 120-125 | 0 | 1 | 1 | 0 | 0 | 0 | 0 |
| 125-130 | 0 | 1 | 2 | 2 | 0 | 1 | 0 |
| 130-135 | 0 | 1 | 0 | 0 | 0 | 0 | 0 |
| 135-140 | 0 | 0 | 0 | 0 | 0 | 2 | 0 |
| 140-145 | 0 | 0 | 0 | 1 | 0 | 1 | 0 |
| 145-150 | 0 | 2 | 0 | 0 | 0 | 0 | 0 |
| 150-155 | 0 | 2 | 0 | 0 | 1 | 1 | 0 |
| 155-160 | 0 | 1 | 1 | 0 | 0 | 0 | 0 |
| 160-165 | 0 | 2 | 0 | 0 | 0 | 0 | 0 |
| 165-170 | 0 | 3 | 0 | 1 | 0 | 0 | 0 |
| 170-175 | 0 | 0 | 0 | 1 | 0 | 0 | 0 |
| 175-180 | 0 | 0 | 1 | 0 | 0 | 0 | 0 |
| 180-185 | 0 | 0 | 0 | 1 | 0 | 0 | 0 |
| 185-190 | 0 | 0 | 0 | 0 | 0 | 1 | 0 |
| 195-200 | 0 | 0 | 0 | 0 | 0 | 1 | 0 |
| 210-215 | 0 | 0 | 0 | 0 | 1 | 0 | 0 |
| 225-230 | 0 | 0 | 1 | 0 | 0 | 0 | 0 |
| 255-260 | 0 | 0 | 1 | 0 | 0 | 0 | 0 |

| Nests area (m²) | $R_1$ | $R_2$ | $R_3$ | $R_4$ | $R_5$ | $R_6$ | $R_7$ |
|---|---|---|---|---|---|---|---|
| 270-275 | 0 | 1 | 0 | 0 | 0 | 0 | 0 |
| >300 | 1 | 0 | 2 | 0 | 1 | 0 | 0 |

**Results and Discussion**

We have applied the iterative algorithm described in the previous section for different values of IB parameters $t$ and $\beta$. The variable $X$ corresponds to nests classes and $Y$ to the regions. The probability $P(x)$ is obtained from table 1 by normalizing the distribution of each nest area. From an initial probability distribution $P(t)$, the conditional probability values $p(t|x)$ are calculated by the iteration process, which proves to be stable with up to 15 iterations to converge. Initial condition testing was performed for each set of parameters $t$ and $\beta$ in order to determine solutions that minimized the functional $L[p(t|x)]$.

For small $\beta$ values, we observe maximum compression where just one cluster is formed regardless of cluster parameter $t$ choice. For the interval $3.5 < \beta < 4.8$ two clusters are formed with the region sets $T_1 : \{R_2, R_4, R_6, R_7\}$ and $T_2 : \{R_1, R_3, R_5\}$. No other cluster is formed for $\beta$ parameter within this interval. Such behavior é obtained from $p(y|t)$ conditional probability which are the same for different values of $t$. This means that for a fixed $\beta$ increase $t$ does not leads to new cluster structures. Above $\beta = 4.8$ three clusters are obtained, $T_1 : \{R_7\}$, $T_2 : \{R_2, R_4, R_6\}$ e $T_3 : \{R_1, R_3, R_5\}$, which are the maximum partition up to $\beta \approx 9.2$.

As a tool to help results analysis and interpretation, we applied a principal component analysis in order to reduce $Y$ dimension size. The first three components are associated with 99.7% explained variance. Eigenvalues and scores for three principal components are presented in table 2.

Table 2. Eigenvalues and scores for three principal components.

|  | Eigenvalues | | |
| --- | --- | --- | --- |
|  | 1 | 2 | 3 |
| Variables | 2053.154 | 88.997 | 7.318 |
| $R_1$ | -0.557759 | 0.285425 | -0.097421 |
| $R_2$ | -0.200217 | -0.220554 | 0.424748 |
| $R_3$ | -0.486617 | 0.217634 | -0.192129 |
| $R_4$ | -0.187064 | -0.392865 | 0.639681 |
| $R_5$ | -0.485939 | 0.111360 | -0.031971 |
| $R_6$ | -0.303610 | -0.077944 | 0.259211 |
| $R_7$ | -0.220711 | -0.806043 | -0.543857 |
| Cumulative | 95.3% | 99.3% | 99.7% |

Fig. 2 and 3 show results of data projection into two and three eigenvectors chosen in eigenvalues descending order.

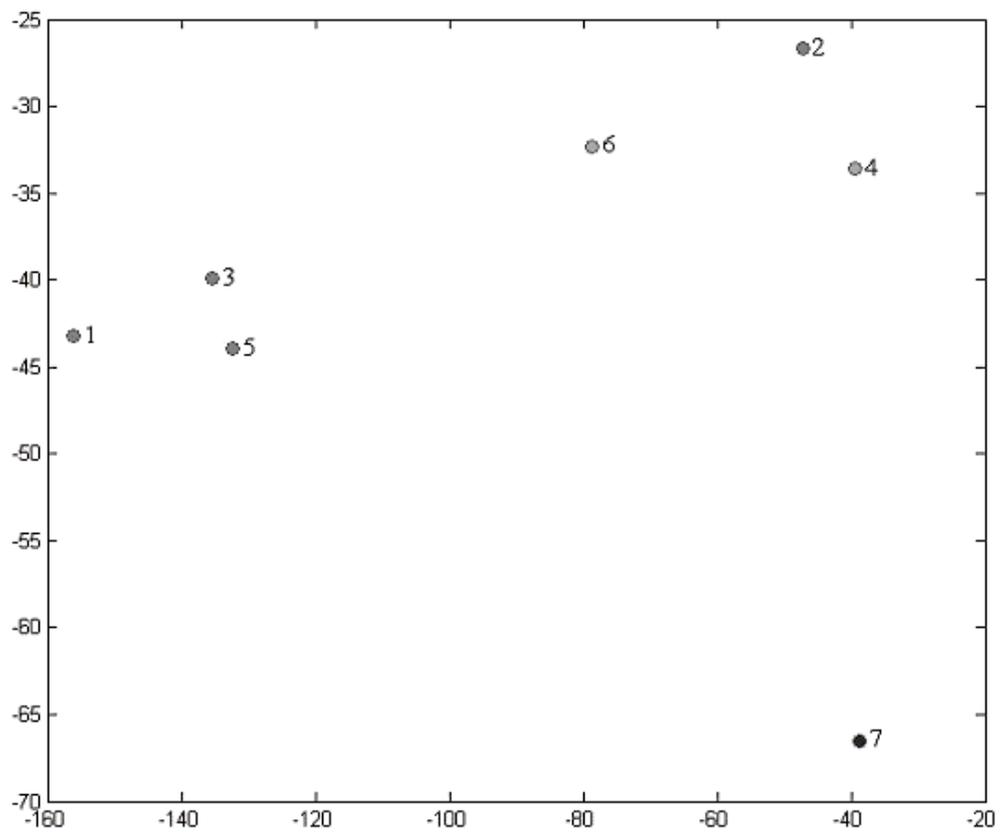

Fig. 2. Data from nests size projected into two eigenvectors with higher eigenvalues.

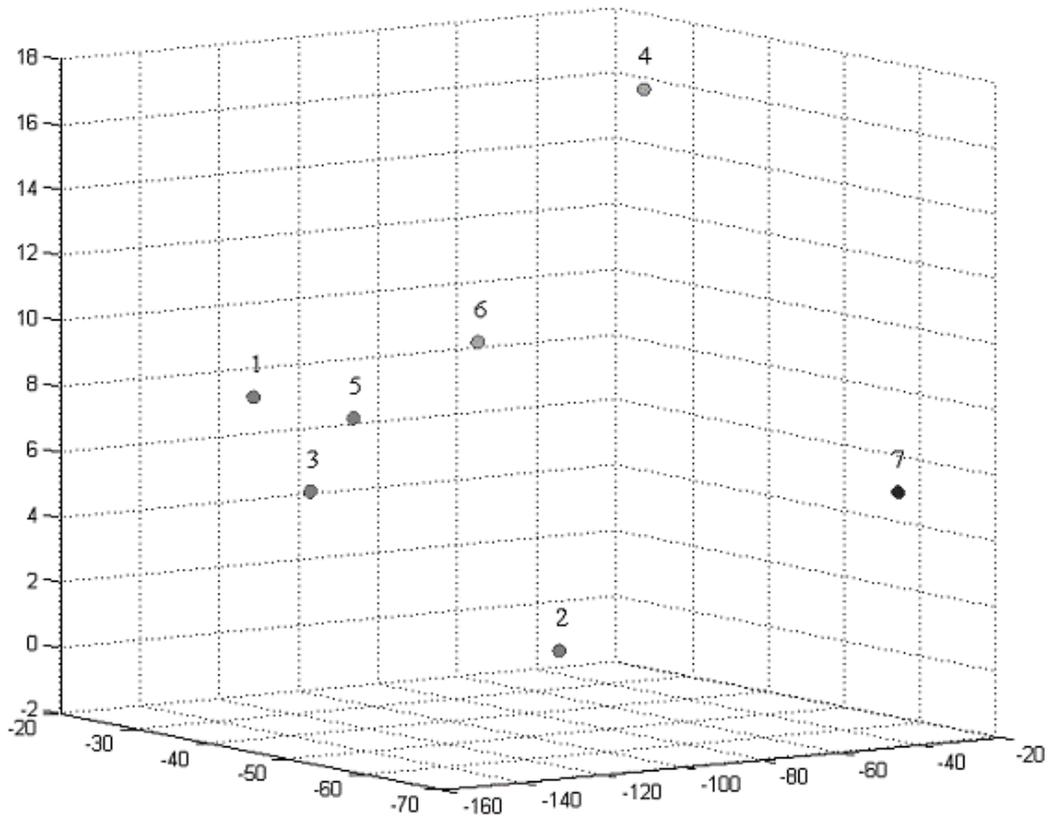

Fig. 3. Data from nests size projected into three eigenvectors with higher eigenvalues.

Observation of Fig.2 suggests two clusters formed by a linear separation of regions $\{R_1, R_3, R_5\}$ from the others. As region 7 is isolated from others of his set, it becomes the first one to be separated into a new cluster formed when $\beta$ is chosen to be higher than 4.8. Although three clusters are observed in the 2D PCA results (Fig.2), it may be of interest to identify four clusters within the data. With IB method and $\beta > 9.2$ the $\{R_2, R_4, R_6\}$ cluster splits into $\{R_4, R_6\}$ and $\{R_2\}$. Within 2D data projection, such separation is not clear, but within the 3D figure (Fig. 3), it may be observed that $R_4$ and $R_6$ are geometrically closer to each other than to $R_2$.

This result is interesting in showing an example in which the PCA dimensionality reduction can lead to a loss of information with respect to clustering. Information Bottleneck method analysis, on the other hand, doesn´t requires prior dimensionality reduction in order to compress data into clusters. This method also allows gradual data compression when parameter $\beta$ is increased, as may be observed in a cluster tree diagram as shown in Fig.4.

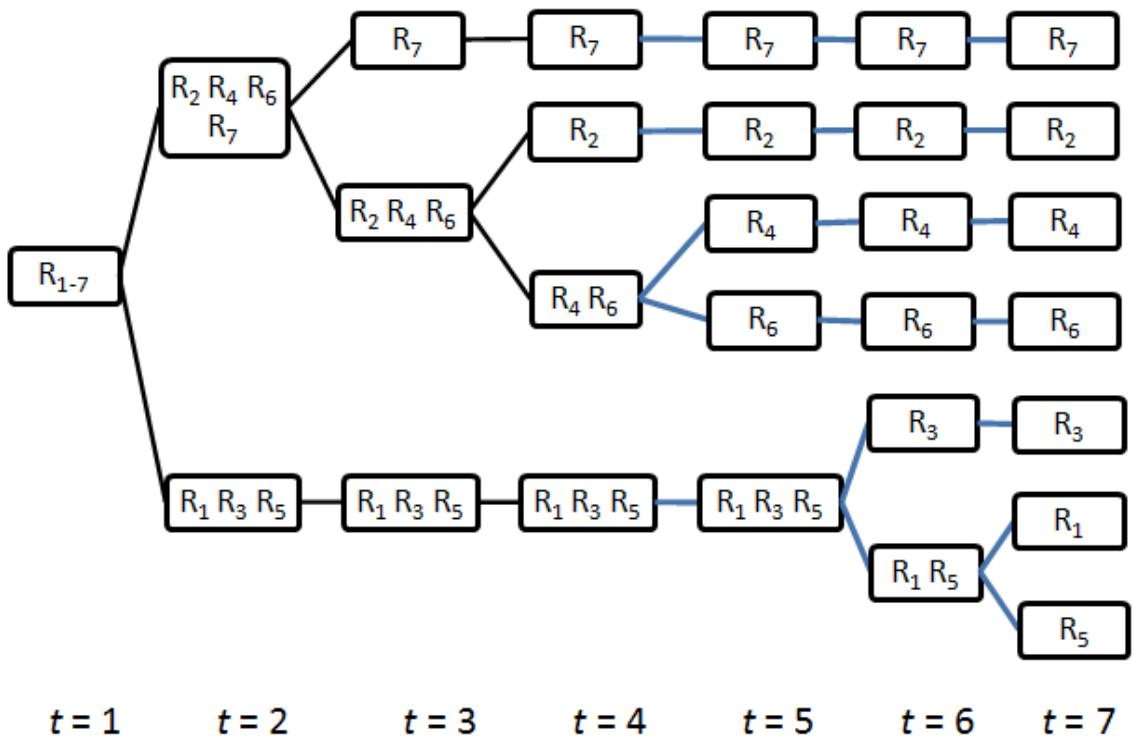

Fig. 4. Cluster tree diagram obtained for different values of parameter $t$ as $\beta \to \infty$. Regions clusters ($Y$) formed using IB method are associated with occurrence of nests areas ($X$), determined from conditional probability $p(x|t)$. Results are shown in Table 2 for the three main clusters.

Regions clusters ($Y$) formed using IB method are associated with occurrence of nests areas ($X$) determined from conditional probability $p(x|t)$. Results are shown in Table 3 for the three main clusters.

Table 3. Ants nests occurrence associated to each cluster by IB method with $t = 3$.

| Nests area (m²) | $R_7$ | $R_2$ | $R_6$ | $R_4$ | $R_1$ | $R_3$ | $R_5$ |
|---|---|---|---|---|---|---|---|
| 5-10 | 41 | 18 | 13 | 24 | 4 | 4 | 10 |
| 10-15 | 32 | 12 | 9 | 16 | 0 | 3 | 5 |
| 15-20 | 26 | 5 | 8 | 16 | 1 | 2 | 3 |
| 20-25 | 15 | 11 | 2 | 2 | 0 | 3 | 3 |
| 25-30 | 7 | 6 | 7 | 9 | 1 | 1 | 2 |
| 35-40 | 0 | 5 | 2 | 9 | 2 | 4 | 1 |
| 50-55 | 0 | 4 | 5 | 5 | 0 | 1 | 0 |
| 55-60 | 0 | 9 | 4 | 3 | 1 | 2 | 0 |
| 70-75 | 0 | 4 | 5 | 1 | 1 | 1 | 1 |
| 80-85 | 0 | 2 | 1 | 0 | 0 | 0 | 0 |

| Nests area (m²) | $R_7$ | $R_2$ | $R_6$ | $R_4$ | $R_1$ | $R_3$ | $R_5$ |
|---|---|---|---|---|---|---|---|
| 110-115 | 0 | 2 | 0 | 1 | 0 | 0 | 0 |
| 115-120 | 0 | 3 | 0 | 0 | 0 | 2 | 0 |
| 125-130 | 0 | 1 | 1 | 2 | 0 | 2 | 0 |
| 130-135 | 0 | 1 | 0 | 0 | 0 | 0 | 0 |
| 145-150 | 0 | 2 | 0 | 0 | 0 | 0 | 0 |
| 160-165 | 0 | 2 | 0 | 0 | 0 | 0 | 0 |
| 165-170 | 0 | 3 | 0 | 1 | 0 | 0 | 0 |
| 270-275 | 0 | 1 | 0 | 0 | 0 | 0 | 0 |
| 30-35 | 0 | 8 | 4 | 9 | 1 | 0 | 4 |
| 40-45 | 0 | 4 | 4 | 5 | 1 | 0 | 1 |
| 45-50 | 0 | 6 | 3 | 9 | 0 | 0 | 3 |
| 60-65 | 0 | 2 | 3 | 4 | 0 | 0 | 0 |
| 85-90 | 0 | 4 | 4 | 2 | 1 | 0 | 1 |
| 90-95 | 0 | 1 | 1 | 2 | 0 | 0 | 0 |
| 95-100 | 0 | 1 | 2 | 3 | 0 | 0 | 0 |
| 100-105 | 0 | 0 | 1 | 1 | 1 | 0 | 0 |
| 105-110 | 0 | 1 | 4 | 0 | 0 | 0 | 0 |
| 135-140 | 0 | 0 | 2 | 0 | 0 | 0 | 0 |
| 140-145 | 0 | 0 | 1 | 1 | 0 | 0 | 0 |
| 150-155 | 0 | 2 | 1 | 0 | 0 | 0 | 1 |
| 170-175 | 0 | 0 | 0 | 1 | 0 | 0 | 0 |
| 180-185 | 0 | 0 | 0 | 1 | 0 | 0 | 0 |
| 185-190 | 0 | 0 | 1 | 0 | 0 | 0 | 0 |
| 195-200 | 0 | 0 | 1 | 0 | 0 | 0 | 0 |
| 0-5 | 62 | 60 | 91 | 55 | 167 | 146 | 145 |
| 65-70 | 0 | 1 | 1 | 1 | 1 | 2 | 1 |
| 75-80 | 0 | 1 | 2 | 0 | 0 | 3 | 0 |
| 120-125 | 0 | 1 | 0 | 0 | 0 | 1 | 0 |
| 155-160 | 0 | 1 | 0 | 0 | 0 | 1 | 0 |
| 175-180 | 0 | 0 | 0 | 0 | 0 | 1 | 0 |
| 210-215 | 0 | 0 | 0 | 0 | 0 | 0 | 1 |
| 225-230 | 0 | 0 | 0 | 0 | 0 | 1 | 0 |
| 255-260 | 0 | 0 | 0 | 0 | 0 | 1 | 0 |
| >300 | 0 | 0 | 0 | 0 | 1 | 2 | 1 |

A misleading interpretation could be performed by administrators, when neighboring regions, for example regions 5 and 6 shown in Fig 1 are considered as belonging to the same cluster. Using IB with PCA on the data enabled a less subjective analysis by providing a precise approach to group the regions for a given number of clusters. For two groups, the IB method shows that regions similar with respect to infestation levels are 2, 4, 6, and 7 in a group, 1, 3 and 5 in another. As parameter cluster $t$ grows, new clusters emerge. The set $\{R_2, R_4, R_6\}$ is dismembered, while there is a significant tendency to preserve $\{R_1, R_3, R_5\}$, which are kept together for $t$ up to 5, as seen in Fig. 4. Region 7 is shown with distinct characteristics from the others for every parameter choice above $t = 2$.

This work shows the application of Information Bottleneck method combined with Principal Component Analysis to group nests of leaf-cutter ants according to their occurrence in each investigated area. While IB allows a hierarchical cluster analysis, PCA furnishes a visual aid by reducing sample dimension.

Procedures used in leaf-cutting ant control, such as granulated baits, are usually based on average size and density of nests. Adequate establishment of such amount avoids waste of resources and help to reduce the environmental impact

The method presented in this work is proposed to aid in the administration of resources by grouping geographic sub-regions that are similar in the distribution of nest sizes (Croft et al. 1983). A possible application is the spatial–temporal investigations of the dissemination of ants around the globe by identifying correlations among different regions (Dillier et al. 2004). It also may be applied to other types of crops and pest infestations, such as coffee leaf miners (Scalon et al. 2011), *Alabama argillacea* (cotton caterpillar) (Fernandes et al. 2003) and larvae of corn (Farias et al. 2001).

**Acknowledgement**

DML Barbato would like to thank Michael Margaliot from Tel Aviv University for insightful discussions on Information Bottleneck method. The authors wish to thank Denis Pretto for collecting the ant nests data.